\documentclass[10pt]{iopart}
\usepackage{bm}
\usepackage{subfigure}
\usepackage{enumerate, pdflscape}
\usepackage{pgfplots,pgf,tikz} \usetikzlibrary{arrows}
\usepackage{booktabs}

\begin{document}
\title[Visual attention while solving the TUG-K]{Visual attention while solving the test of understanding graphs in kinematics: An eye-tracking analysis}
\author{P Klein$^1$, A  Lichtenberger$^2$, S K\"{u}chemann$^1$, S Becker$^1$, M~Kekule$^3$, J Viiri$^4$, C Baadte$^1$, A  Vaterlaus$^2$ and J Kuhn$^1$}
\address{$^1$Department of Physics, Physics Education Research, Technische Universit\"{a}t Kaiserslautern, Erwin-Schr\"odinger-Str. 46, 67663 Kaiserslautern, Germany}
\address{$^2$Solid-State Dynamics and Education, Department of Physics, ETH Z\"urich, John-von-Neumann-Weg 9, 8093 Z\"urich, Switzerland}
\address{$^3$Department of Physics Education, Faculty of Mathematics and Physics, Charles University, V Holesovickach, 180 00 Praque, Czech Republic}
\address{$^4$Department of Teacher Education, University of Jyv\"askyl\"a, 40014 Jyv\"askyl\"a, Finland}
\ead{pklein@physik.uni-kl.de}
\bibliographystyle{unsrt}
\begin{abstract}
This study used eye-tracking to capture the students' visual attention while taking the test of understanding graphs in kinematics (TUG-K). A total of $N=115$ upper-secondary-level students from Germany and Switzerland took the 26-item multiple-choice instrument after learning about kinematics graphs in the regular classroom.  Besides choosing the correct alternative among research-based distractors, the students were required to judge their response confidence for each question. The items were presented sequentially on a computer screen equipped with a remote eye tracker, resulting in a set of approx. 3000 paired responses (accuracy and confidence) and about  40 hours of eye movement-data (approx. 500.000 fixations). The analysis of students' visual attention related to the item stems (questions) and the item options reveal that high response confidence is correlated with shorter visit duration on both elements of the items. While the students' response accuracy and their response confidence are highly correlated on the score level, $r(115)=0.63, p<0.001$, the eye-tracking measures do not sufficiently discriminate between correct and incorrect responses. However, a more fine-grained analysis of visual attention based on different answer options reveals a significant discrimination between correct and incorrect answers in terms of an interaction effect: Incorrect responses are associated with longer visit durations on strong distractors and less time spent  on correct options while correct responses show the opposite trend. Outcomes of this study provide new insights into the validation of concept inventories based on students' behavioural level.

\end{abstract}

%%
%% Uncomment for keywords
%\noindent{\it Keywords}: eye tracking, TUG-K, kinematics graphs, confidence assessment, instrument validation

\submitto{\EJP}
\maketitle

\section{Introduction}
In 1994, Robert Beichner introduced the test of understanding graphs in kinematics (TUG-K) to the physics education research community \cite{Bei94}, which has become one of the most widely used test to date designed to evaluate students' understanding in this subject (see, for example, Refs. \cite{Perez10, MS13, Gurel15}). Since then, the TUG-K was used by teachers and researchers to assess students' understanding of graphs in kinematics, as well as their learning about them. For example, the TUG-K has been used as a pre- and posttest to investigate the effectiveness of instructions (e.g., video-based motion analysis \cite{Bei96}) or to study the relationship between the understanding of kinematics graphs and other variables (e.g., gender) \cite{Bektasli12}. The test has also been used as a reference to design new, related tests: for example, a test to assess the understanding of graphs in the context of calculus \cite{Perez10}, the kinematics concept test \cite{Lichtenberger} or the kinematics representational competence inventory \cite{KIRC}. For the TUG-K, the graphs of all test items relate to motion in one dimension and the items address the concepts of graph slope and area under the curve for different objectives  \cite{Bei94}. All items were created based on extensive  research on student difficulties with graphs of position, velocity, and acceleration versus time. Besides the correct alternative, the items offer incorrect options (\textit{distractors}) that address typical  student difficulties with gaphs, such as the graph-as-picture error \cite{Bei94}.  

Recently, several modifications of the TUG-K have been made to achieve parallelism in the objectives of the test \cite{Zavala}. The modified version of the test was proven to fulfil the statistical tests of difficulty, discriminatory power, and reliability; also, that the great majority of the modified distractors were effective in terms of their frequency selection and discriminatory power.  Although much research has been done to validate the instrument properties,  the methods being used are limited to classical test analysis and students interviews \cite{Lindell, Adams}. While analysing written responses and student interviews is very effective for identifying students misconceptions and for detecting typical sources of erroneous reasoning, these methods are very time-consuming and require interactions between an interviewer and the interviewee  \cite{Polking07}. Consequently, these methods are typically used in an early stage of test construction. In addition, the data gathered in large-scale assessment scenarios described above are typically reduced to the frequency of distractors and the response accuracy rates, neglecting the capability to directly probe the students' thinking processes at a finer level during problem solving.  One way to tackle these problems could be the analyses of gaze data during test completion by eye-tracking technology. 

Meanwhile, eye tracking has received attention from the physics education research community. Previous eye-tracking research on students' understanding of kinematic graphs revealed that students with low spatial abilities tend to interpret graphs literally \cite{Kozhevnikov}, yielding a link between an important cognitive variable (visuo-spatial abilities) and graph understanding.  In the similar context, Madsen \textit{et al.} showed that students who answered a question correctly spent more time on relevant areas of a graph, e.g., the axes \cite{Madsen}.  Their finding also suggests a link between the previous exposure to a type of a problem and the ability to focus on important regions \cite{Madsen2}.  Recently, Susac \textit{et al.} compared students' understanding of graphs between physics and non-physics students in different contexts (viz., physics and finance) \cite{Susac}. The authors found that physics students outperformed the non-physics students for both contexts. The analysis of eye-movement data revealed that physics students focused significantly longer on the graph and spent more time looking at questions in the unfamiliar context of finance. The results have broadly been confirmed by Klein \textit{et al.} in a replication study using physics students and a different non-physics sample, viz. economics students \cite{Klein2019}. The physics students solved the problems also better than the non-physics students but, in contrast to the work by Susac \textit{et al.}, Klein \textit{et al.} found that both groups of students had very similar visit durations on the graphs, consequently proving total visit duration to be an inadequate predictor of performance. These contradictory results obtained in the studies require further investigations.  

Some items of the TUG-K have been investigated with eye-tracking technology before. Kekule reported qualitative results that indicate different task-solving approaches between best-and worst-performing students when they solved some specific items of the TUG-K \cite{Martina1, Martina2}. However, up to now there is no study that used the complete test and applied a rigorous analysis to all test items. Given that the TUG-K is  continuously applied for research purposes and researchers  are still putting effort in its development and modification, a throughout analysis of visual attention might reveal further information about test validity, distractor functionality, discrimination, and about the  students' cognitive processes involved in problem solving on the test level. Accurate measures on which distractors students spend time looking or how they distribute their attention among the questions and the answer options will provide insights on the cognitive demands of the complete test rather than individual items on a behavioural level.

Following this line of research, Han \textit{et al.} recently used eye-tracking technology to investigate students' visual attention while taking the complete Force Concept Inventory (FCI) in a web-based interface \cite{Han}. They compared two samples of students, one that was tested before instruction and the other after attending classical mechanics lectures for several weeks. The authors were able to show that the students' performance increases but there was no correlation between the performance and the time spent to complete the test. The authors also investigated three questions about Newton's third law more carefully. They found that even though the students' expertise shifted towards more expert like thinking, the students still kept a high level of attention at incorrect choices that addressed misconceptions, indicating significant conceptual mixing and competition during problem solving while answering these three items. In contrast to the TUG-K, the FCI does not emphasize a special kind of representation (i.e., graphs) and uses exclusively text-based options. Hence, there is value in expanding the research methodology to investigating complete item sets of other concept inventories. In addition, it is also of importance to extend the method of classifying answer choices to all test items (instead of using a subset) to generalize the hypothesis of conceptual mixing and to investigate how the mixing is influenced by students expertise. 

Apart from eye tracking, the assessment of confidence ratings adds more information about the quality of students' understanding when answering multiple choice questions \cite{Aslanides2013, Hasan,Lindsey,Dowd}. In general, confidence judgements reflect the examinee's belief in the correctness of the chosen alternative  and can be interpreted as one aspect of metacognition \cite{Aslanides2013}. Being able to distinguish between right and wrong answers may be a condition to reflect and regulate learning and is linked to the students' ability to evaluate their own understanding. In the above mentioned eye-tracking study on students graphical understanding of the slope and the area concept, Klein \textit{et al.} assessed the response confidence level of the students and found that physics students have a higher ability to correctly judge their own performance compared to economics students \cite{Klein2019}. While the authors evaluated the confidence scores quantitatively, they did not link the confidence results to the eye-tracking data they obtained, leaving this question open for future research. In an investigation about a more advanced physics topic (viz., rotational frames of reference), K\"{u}chemann \textit{et al.} used two multiple choice questions to assess students accuracy and  response confidence after the students observed real experiments about non-inertial frames of references \cite{Kuechemann}. The authors identified that longer visit durations were related to low confidence ratings, suggesting that students with low confidence  lack a connectional understanding of the representations involved and tried longer to make referential connections between them as compared to confident students. 

\section{Research questions} \label{sec:RQ}
In this study, eye tracking was used to capture students' visual attention while completing a computer-based version of the  TUG-K. The goal of the study was to investigate how the students' conceptual understanding of kinematics graphs influenced the students' allocation of attention to the questions and to the different answer options. Additionally, the study explored how the students' response confidence was related to their eye movements and to their conceptual understanding. To ensure that the students had sufficiently prior knowledge to reasonable understand all questions, the test was administered as a post test after the topic was taught at school. The research questions are twofold:

\begin{enumerate}[1.]
\item How is the time spent on individual questions and options related to  students response accuracy and to students' response confidence?
%\item How is the transitional behaviour of students on individual questions related to the outcome measures?
\item How do students of different expertise (as indicated by the total test score) distribute their attention on different answer options? 

\end{enumerate}

For the second research question, the different option choices of individual items were classified based on a distractor analysis, resulting in four categories: correct option, most popular incorrect options (oftentimes addressing misconceptions), popular incorrect options and unpopular (implausible) incorrect options options, see Section \ref{sec:materials} for details. Based on prior findings of expertise research \cite{Geg11}, the following hypothesis was examined:

\begin{enumerate}[H]
\item Experts spend more time on conceptually relevant areas of the option (i.e., the correct alternatives) and less time on conceptually irrelevant areas (i.e., the strong distractors and the implausible options) compared to novices.  
\end{enumerate}

\section{Method}
\subsection{Subjects and data collection}
The sample consisted of upper-secondary students from two German grammar schools (``Gymnasium", $N=68$) and one Swiss Gymnasium ($N=47$), including 12 different physics courses in total.  When the test was administered, the subject kinematics was completed in all courses. As stated by the  physics teachers, the formal basics of kinematics have been taught up to this point and students have learned already about different types of motion, including demonstration experiments. Therefore we can consider the measurement in this study as a post-test. All physics teachers discussed about kinematics graphs in their courses but the intensity was neither assessed nor controlled. Apart from that, no special educational requirements were necessary with regard to the participating physics courses. 

All students  (58 female, 57 male; all with normal or correct-to-normal vision) completed the TUG-K in its original sequence (26 questions in German language) on a computer screen equipped with an eye tracker. Four identical eye-tracking systems were set up in the school libraries and the students took part in the experiment  in groups of up to four either during their spare time or during regular lessons (given that they had the teachers' permission).   At the beginning of the experiment, students were placed in front of a computer screen, and a nine-point calibration and validation procedure was used.  A researcher (one of the co-authors) instructed the students before the experiment started, carried out the eye-tracking calibration, and checked the accuracy of the gaze detection during the experiment.  Besides the test takers and the researchers, no other persons were present in the room. % and noise cancelling headphones were offered to the students for better concentrating  on the visual stimuli. 
The students read the material without any interruptions by the researcher. The students were free to use as much time as needed for answering the questions. Whenever a student was ready to give an answer, they pressed a button and gave their answer as well as their response confidence. The students did not receive any feedback after completing a task and were unable to skip back to previous tasks.

\subsection{Materials}\label{sec:materials}

\begin{figure}[t!]
\centering
\fbox{\includegraphics[width=0.8\linewidth]{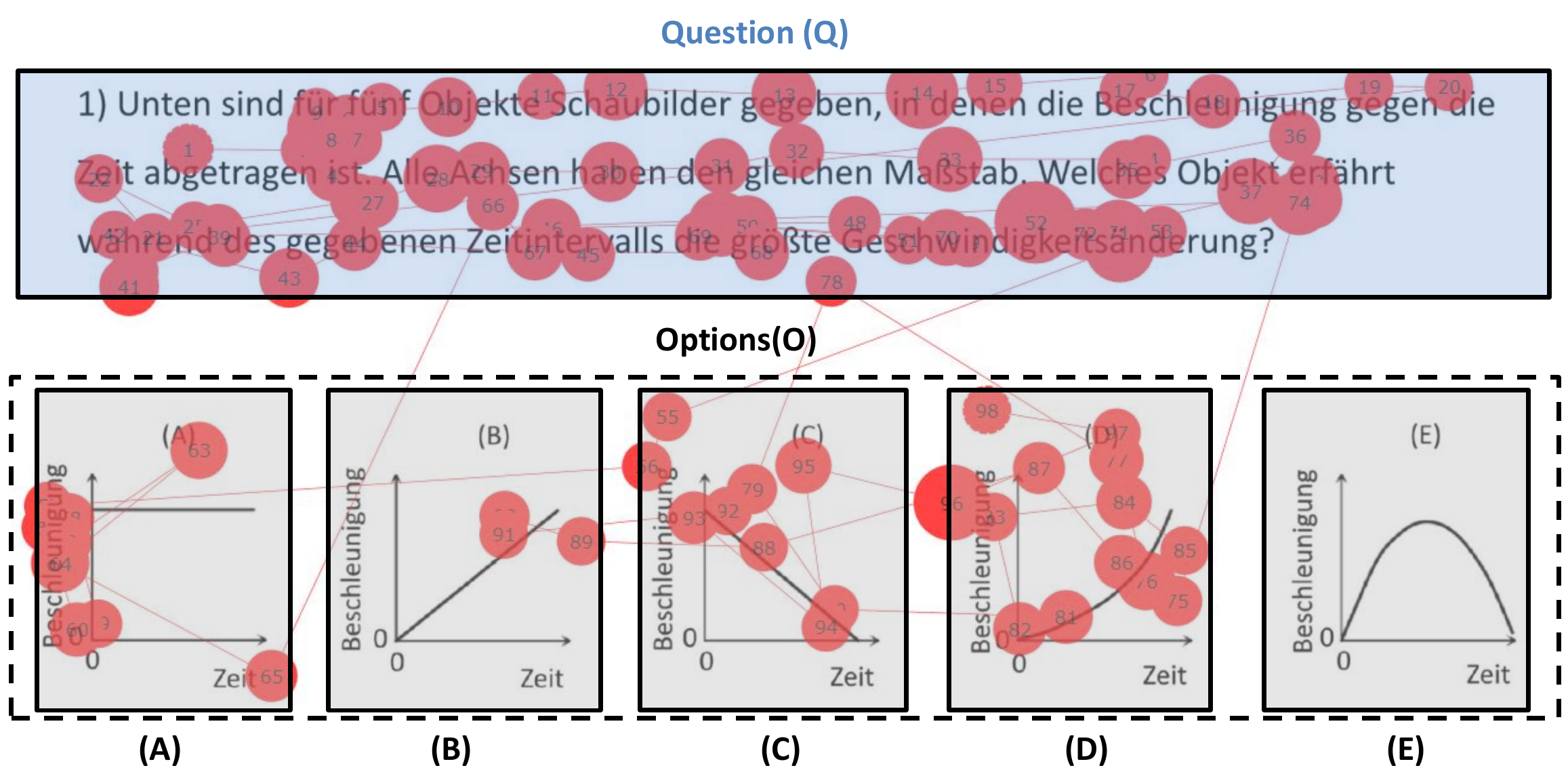}}
\caption{Definition of areas of interest for the TUG-K item 1 with an excerpt of an individual gaze plot. Regular eye movement alternates between fixations (red circles, centered at the position of fixation) and saccades (lines between fixations). Increasing numbers reflect the chronological order of fixations, and the size of the circles illustrates the duration of the fixation. }
\label{fig:AOIs}
\end{figure}

The TUG-K contains 26 multiple-choice questions. The original version of the test exists in German language \cite{physport} but the modified version (TUG-K 4.0)---which was used in this study---does not. Therefore, the modified items were translated to German.%\footnote{The material is available here: Set Link}. 
The test was converted into a computerized assessment using the Tobii Studio eye-tracking software which was also used to present the stimulus to the subjects and to host the eye tracker \cite{tobii}. Each item was presented on one single slide, starting with the item question on the top of the page. The item question contained either only text or text and a graph. Each item had five answer options presented either as graphs or as text (written words or numbers). The concepts addressed in the single items can be found in the original publication of the test \cite{Bei94, Zavala} and a short overview is given in Table \ref{tab:content} in the appendix. The TUG-K is a research based instrument, and all items are developed based on qualitative studies that pointed towards typical student errors and misconceptions with kinematic graphs. In many cases, the incorrect answer options are well-designed to address typical  learning difficulties with kinematic graphs. These strong distractors were constructed based on empirical research and typically receive many votes from students who answer incorrectly.  However, some of the alternatives are quite implausible and do not reflect popular misconceptions. These alternatives were only selected by a minority of students who answer incorrectly as indicated in prior investigations \cite{Zavala} and by our own data set.  Based on the test takers choices of the distractors provided in the original work ($N\approx 500$) \cite{Zavala}, the options were split into four categories: 
\begin{itemize}
\item correct option 
\item most popular incorrect  options (highest proportion of incorrect choices)
\item popular incorrect options (more than 10\% and less than the most popular choices)
\item unpopular incorrect options (quite implausible and only few choices ($\leq 10 \%$))
\end{itemize}
An overview about the categorization is provided in the appendix.  Two areas of interest (AOIs) were defined for each item, covering the question (text or text only and graph; \textit{``Q"}) and all five options (\textit{``O"}). Furthermore, five smaller rectangular AOIs (\textit{``A"}--\textit{``E"}), equally in size and non-overlapping, were defined for every item, each surrounding one of the five answer options. \textit{Q} therefore contains the subsets (\textit{A}--\textit{E}) and some white space between the options. Figure \ref{fig:AOIs} shows one example for the definition of AOIs with an excerpt of the eye-tracking data from one student. Since we do not compare the items among each other, there is no effect if the AOIs are different in size among different questions. Based on the measurement data, 96\% of the fixations were located inside  $Q \cup O$. 

\subsection{Eye-tracking apparatus}
The tasks were presented on a 22-in. computer screen. The resolution of the computer screen was set to 1920 $\times$ 1080 pixels with a refresh rate of
75\,Hz. Eye movements were recorded with a Tobii X3-120 stationary eye-tracking system \cite{tobii}, which had an accuracy of less than 0.40$^\circ$ of visual angle (as reported by the manufacturer) and a sampling frequency of 120\,Hz. The system allows a relatively high degree of freedom in terms of head movement (no chin rest was used). To detect fixations and saccades, an I-VT (Identification by Velocity Threshold) algorithm was adopted \cite{salvucci2000identifying}. An eye movement was classified as a saccade (i.e., in motion) if the acceleration of the eyes exceeded 8500$^\circ/$s$^2$ and velocity exceeded 30$^\circ/$s.

%The results are based on the AOIs described above, and averages are computed for every participant and for each AOI per item. The dimensions of the Question and Option AOIs are larger than 150 mm $\times$ 70 mm when displayed on the monitor, which give an approximate angular dimension of 14$^\circ$ $\times$ 7$^\circ$. Therefore, the measurement uncertainty of 0.40$^\circ$ is sufficient to separate fixations inside or outside the AOIs. In all questions, the distance between the boundaries of AOIs covering single answer options was at least 20 mm ($\equiv$ 2$^\circ$ visual angle). 

\subsection{Data and data analysis} \label{sec:analysis}
For the analysis we used the \textit{response accuracy }(0 = incorrect response, 1 = correct response), the \textit{response confidence} (as measured by a Likert scale ranging from 1 to 6), and the  \textit{visit duration}. The confidence ratings were linearly transformed to a $[0,1]$-scale where 0 means lowest and 1 means highest confidence. The \textit{confidence index} and the \textit{difficulty index} then refer to the average of the students' confidence ratings and accuracy scores per item, respectively. The median per item was used to define three confidence levels: low (below median), intermediate (equals median), and high (beyond median). The visit duration was calculated for the single AOIs (\textit{Q,O, A--E)}. This measure was conveniently  extracted using the Tobii Studio software.
%\item \textit{Number of Q--O and O--O transitions}. To calculate the number of transitions between \textit{O} and \textit{Q}, all data were exported through a spreadsheet, and the saccades from \textit{O}  to \textit{Q} and vice versa were counted numerically. It was not distinguished between \textit{Q}--\textit{O} and \textit{O}--\textit{Q} transitions. The same procedure was applied to determine the transitions between the different answer options \textit{A}--\textit{E}, yielding ${5 \choose 2} = 20$ possibilities.   

A two-factorial analysis of variance (two-way ANOVA) was used to investigate the differences in the eye-tracking measures between response accuracy levels (factor 1; correct vs. incorrect) and between response confidence levels (factor 2; high vs. intermediate vs. low) as well as their interaction. Given the huge dataset (26 items $\times$ 115 students $\approx$ 3000 data points), each effect was considered statistical significant when the $p$-value was below the 0.1\% threshold ($p<0.001$). In these cases, we also report the effect size measure in terms of Cohen's $d$ to judge the magnitude of the phenomenon. Descriptors of magnitudes for $d = 0.01$ to 2.0 have initially been suggested  by Cohen and expanded by Sawilowsky; $d>0.01$: very small, $d>0.20$: small, $d>0.50$: medium, $d>0.80$: large, $d>1.20$: very large, $d>2.0$: huge \cite{Cohen, Sawil09}.

Furthermore, the students' visit duration on options representing different concept models  were compared using a repeated measure ANOVA with the attention on option choices as the within-subject variable (correct vs. incorrect options or more fine-grained categories, see above) and the expertise level as the between-subject variable (20\%-percentiles based on the students' total test score).  

\section{Results}
%First we report about the students' mean scores and the item statistics (i.e., on the macro level, Sec. \ref{sec:macro}). After that, the data is analysed with respect to the research questions 1 and 2 that were formulated in Sec. \ref{sec:RQ}. Finally,  the research hypothesis is investigated using a more detailed analysis of the visual attention on different answer options.  
\subsection{Descriptives and correlations} \label{sec:macro}
The students' mean test score and standard deviation were (59$\pm$25\%), ranging from 4\% (1 student) to 100\% (2 students). The mean response confidence score was (72$\pm 15$)\%, ranging from  29\% (1 student) to 100\% (1 student). Correct answers ($N=1773$) were given with higher response confidence (78\%) than incorrect answers ($N=1217$; 63\%), and consequently, the students' mean response confidence score for correct answers (74\%) was higher compared to incorrect answers (67\%), $t(112)=6.6, p<0.001$.  Correlation analysis also confirmed the result, which showed that  accuracy and confidence scores are significantly correlated, $r(115)=0.63, p<0.001$.

Table \ref{tab:items} summarizes the students' response accuracy and confidence scores for each item. The items cover a reasonable range of difficulty from 0.34 (hardest; item 1) to 0.89 (easiest; item 13 and item 14), and the averaged index of 0.59 falls into the suggested range by \cite{DB09}. The confidence index varies from 0.57 (item 10) to 0.83 (item 25) with average 0.72. Both indices are correlated on the item level, $r(26)=0.57, p<0.01$.   

 The eye-tracking data was used to calculate the time required to complete the TUG-K, the time spent on each single TUG-K question, and the visit duration on different elements of each question. The average time spent completing the TUG-K was (20.2$\pm$4.6) min, ranging from 10.4 to 34.1 min. There is no significant correlation between total time spent on the TUG-K and the accuracy score, $r(115)=-0.17, p>0.05$, whereas the correlation between time spent on the test and confidence scores is significant, $r(115)=-0.25, p<0.01$.

%\begin{landscape}
\begin{table}[h!]
\centering \footnotesize
\caption{Item characteristics.}
\label{tab:items}
\begin{tabular}{ccc cc c c c c cc c cc} \hline\hline
 \addlinespace[3pt]
 & Difficulty & Confidence && \multicolumn{5}{c}{Percentage} && \multicolumn{2}{c}{Visit Duration (sec)} %&& \multicolumn{2}{c}{Number of Transitions}
 \\ \cmidrule{5-9}   \cmidrule{11-12} %\cmidrule{14-15} 
Item	&	 Index				&	 Index				&&	A	&	B	&	C	&	D	&	E	&&	Question 			&	Options 				%&&	\textit{Q}--\textit{O}				&	\textit{O}--\textit{O}				
\\ \hline  \addlinespace[3pt]
1	&	0.34	(0.04)	&	0.68	(0.06)	&&	34	&	12	&	1	&	13	&	40	&&	16.8	(0.6)	&	24.1	(1.2)	\\
2	&	0.81	(0.04)	&	0.81	(0.08)	&&	2	&	5	&	8	&	4	&	81	&&	29.5	(1.3	)	&	4.1	(0.2)	\\
3	&	0.81	(0.04)	&	0.77	(0.07)	&&	2	&	1	&	13	&	81	&	3	&&	17.6	(0.9)	&	22.3	(1.0)	\\
4	&	0.38	(0.05)	&	0.70	(0.07)	&&	2	&	10	&	15	&	38	&	36	&&	56.4	(2.4)	&	6.7	(0.5)	\\
5	&	0.53	(0.05)	&	0.81	(0.08)	&&	3	&	0	&	53	&	33	&	10	&&	33.8	(1.5)	&	4.1	(0.3)	\\
6	&	0.62	(0.05)	&	0.59	(0.06)	&&	11	&	62	&	10	&	11	&	6	&&	48.0	(2.2)	&	10.7	(0.6)	\\
7	&	0.50	(0.05)	&	0.63	(0.06)	&&	50	&	20	&	12	&	14	&	3	&&	48.2	(2.2)	&	8.0	(0.6)	\\
8	&	0.74	(0.04)	&	0.77	(0.07)	&&	2	&	15	&	7	&	74	&	3	&&	20.2	(0.9)	&	29.1	(1.3)	\\
9	&	0.56	(0.05)	&	0.77	(0.07)	&&	11	&	23	&	6	&	3	&	56	&&	17.2	(0.7)	&	19.7	(1.1)	\\
10	&	0.44	(0.05)	&	0.57	(0.05)	&&	44	&	30	&	6	&	8	&	12	&&	40.6	(1.7)	&	24.8	(1.0)	\\
11	&	0.63	(0.05)	&	0.75	(0.07)	&&	4	&	22	&	6	&	63	&	4	&&	27.0	(1.0)	&	19.9	(1.0)	\\
12	&	0.56	(0.05)	&	0.81	(0.08)	&&	6	&	56	&	6	&	14	&	18	&&	10.8	(0.5)	&	23.5	(1.3)	\\
13	&	0.89	(0.03)	&	0.79	(0.07)	&&	89	&	7	&	0	&	3	&	2	&&	24.0	(1.1)	&	2.6	(0.2)	\\
14	&	0.89	(0.03)	&	0.75	(0.07)	&&	3	&	89	&	2	&	2	&	4	&&	30.4	(1.3)	&	20.1	(1.5)	\\
15	&	0.56	(0.05)	&	0.69	(0.06)	&&	56	&	5	&	21	&	1	&	17	&&	30.5	(1.4)	&	28.9	(1.9)	\\
16	&	0.37	(0.05)	&	0.61	(0.06)	&&	8	&	30	&	13	&	37	&	13	&&	34.0	(1.8)	&	5.9	(0.4)	\\
17	&	0.68	(0.04)	&	0.71	(0.07)	&&	68	&	12	&	2	&	18	&	0	&&	12.2	(0.7)	&	26.9	(1.3)	\\
18	&	0.53	(0.05)	&	0.65	(0.06)	&&	53	&	16	&	5	&	20	&	6	&&	41.2	(2.1)	&	8.4	(0.7)	\\
19	&	0.43	(0.05)	&	0.60	(0.06)	&&	4	&	43	&	20	&	29	&	3	&&	28.2	(1.4)	&	18.1	(1.0)	\\
20	&	0.47	(0.05)	&	0.77	(0.07)	&&	2	&	22	&	47	&	15	&	15	&&	10.0	(0.5)	&	20.6	(1.1)	\\
21	&	0.64	(0.04)	&	0.73	(0.07)	&&	2	&	64	&	3	&	14	&	17	&&	26.2	(1.0)	&	19.6	(0.9)	\\
22	&	0.62	(0.05)	&	0.70	(0.07)	&&	8	&	9	&	62	&	10	&	12	&&	16.3	(0.7)	&	18.3	(1.0)	\\
23	&	0.55	(0.05)	&	0.68	(0.06)	&&	0	&	55	&	6	&	25	&	14	&&	14.9	(0.8)	&	20.9	(1.1)	\\
24	&	0.51	(0.05)	&	0.71	(0.07)	&&	51	&	35	&	10	&	4	&	0	&&	14.7	(0.7)	&	20.2	(1.0)	\\
25	&	0.70	(0.04)	&	0.83	(0.08)	&&	7	&	4	&	70	&	18	&	1	&&	10.4	(0.8)	&	12.6	(0.5)	\\
26	&	0.67	(0.04)	&	0.75	(0.07)	&&	3	&	67	&	1	&	18	&	11	&&	10.2	(0.5)	&	16.3	(1.2)	\\

 \hline\hline

\end{tabular}
\end{table}

%\end{landscape}

For each item, the mean visit duration was split into time spent on the question (\textit{Q}) and on time spent on the options (\textit{O}), see Table \ref{tab:items}. On the item level, there is no correlation between the difficulty index and any of the visit-duration measures, confirming the result from above. On the contrary, the confidence index is correlated with the visit duration on the question $r(26)=-0.54, p<0.01$ but not with time spent on the options.  %Furthermore, the eye-tracking data was used to determine the number of transitions between relevant elements of an item as described in Sect. \ref{sec:analysis}. We distinguish between the mean number of transitions between the question and the options (\textit{Q}--\textit{O} transitions) and between the mean number of transitions between different answer options (\textit{O}--\textit{O} transitions). No correlation between these measures and the difficulty/confidence index was found on the item level.

\subsection{Visit duration on questions and options}

\begin{figure}
\centering
\includegraphics[width=\linewidth]{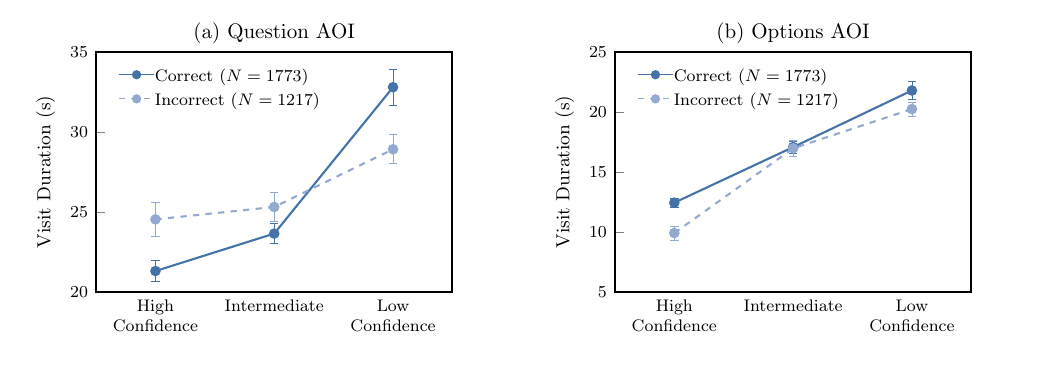}
\caption{Students' mean visit duration for (a) questions and (b) options. %(c, d) Mean number of transitions between question and options and between different options. 
The eye-tracking measures are summarized for correct and incorrect responses by the students' confidence level.  The error bars reflect the standard errors.}
\label{fig:acc_conf}
\end{figure}

\begin{table}[b]
\centering \small
\caption{Results of the univariate ANOVAs conducted on four eye-tracking measures (VD = Visit Duration, Q = Question, O = Options) with accuracy (correct vs. incorrect) as factor 1 and confidence (low vs. intermediate vs. high) as factor 2. For each measure, the analysis was based on the complete data set.}
\label{tab:anova}
\begin{tabular}{lcccccccc} \hline\hline  
 \addlinespace[3pt]
 & \multicolumn{2}{c}{Accuracy $(df=1)$} && \multicolumn{2}{c}{Confidence $(df=2)$} && \multicolumn{2}{c}{Interaction $(df=2)$} \\  \cmidrule{2-3}  \cmidrule{5-6} \cmidrule{8-9}  
 &~~~~~~$F$~~~~~~&~~~~~~$d$~~~~~~&&~~~~~~$F$~~~~~~&~~~~~~$d$~~~~~~&&~~~~~~$F$~~~~~~&~~~~~~$d$~~~~~~ \\ \hline
  \addlinespace[3pt]
VD (Q)	&	0.21		&	... 	&&	44.62$^{*}$	&	0.20	&&	8.75$^{*}$	&	0.07	\\
VD (O) &	7.59	&	... 	&&	118.20$^{*}$	&	0.34	&&	1.91	&	...	\\
%NT (Q-O) && 5.88 & ... && 84.55$^{*}$ & 0.29 && 0.39 & ...\\
%NT (O-O) && 0.00 & ... && 80.16 & 0.29 && 5.65 & ... \\
\hline\hline 
 \multicolumn{8}{l}{$^{*}p<0.001$}
 \end{tabular}

\end{table}
Student's time spent on the different elements of an item (question or options) was analysed using a two-way ANOVA with response accuracy as factor 1, confidence as factor 2, and the time as the dependent variable. The results are summarized in Fig. \ref{fig:acc_conf} (a,b) and Table \ref{tab:anova}. For visit duration on the question, we found a statistically significant main effect of confidence [$F(2, 2984)=44.6, p\ll 10^{-20}$, small effect size $d=0.20$], whereas the response accuracy had no significant impact [$F(1, 2984)=0.2, p>0.001$]. Figure \ref{fig:acc_conf} (a) shows that low confidence is related to longer visit durations both for correct and for incorrect responses. However, the difference between low and intermediate confidence is bigger for correct responses than for incorrect responses, indicating an interaction effect. Indeed, the interaction between confidence and accuracy is statistically significant but the size of the effect is negligible  [$F(2, 2984)=8.75, p=0.001$, very small effect size $d=0.07$]. 

For visit duration on the options we found a significant main effect of confidence with small effect size ($d=0.34$) whereas the accuracy had no impact on the visit duration on the options. Figure \ref{fig:acc_conf} (b) again shows that low confidence is related to longer visit durations, and the trend is very similar between correct and incorrect responses without an interaction effect.

%To analyse the number of transitions that the students performed while solving the single test items, we applied the same analysis procedure that we applied to the visit durations.  For both types of transitions (\textit{Q-O} and \textit{O-O}), we found a main effect of the factor  confidence with small effect sizes: $F(2, 2984)=84.55, p\ll10^{-20}, d=0.29$ for \textit{Q-O} transitions and $F(2, 2984)=80.16, p\ll10^{-20}, d=0.29$ for \textit{O-O} transitions. The Figs. \ref{fig:acc_conf} (a) and (b) reveal that responses given with high confidence had a lower number of transitions both between \textit{Q} and \textit{O} and between the different options than responses given with intermediate or low confidence. On the contrary, response accuracy had no significant impact on the number of transitions and there was also no interaction between both measures.

\subsection{Students' visual attention on different answer options}
Student's fixations on different answer options reveal information about their reasoning process when solving problems on kinematics graphs. As described in Sec. \ref{sec:materials}, most of the incorrect alternatives were constructed based on student difficulties with graphs and prior studies showed that some of them were more popular than others (in terms of being selected by the students), see Table \ref{tab:options} in the Appendix. Analysis of visual attention on different answer options might therefore reveal important information about the attractiveness of the alternatives and the presence of misconceptions on a behavioural level. The results in the previous section showed that the students' understanding---as reflected by their response accuracy (i.e., correct or incorrect)---is not associated with the time spent on the options. %This suggests a similar amount of cognitive effort to solve the questions, independent on answering correctly or incorrectly. 
As a next step, the answer options were distinguished more carefully: (i) Visual attention on correct and incorrect options is investigated and related to the response accuracy and (ii) three categories of incorrect options are differentiated (most popular, popular, and unpopular) and are being related to the students' expertise level as an indicator of their conceptual states. For the  latter purpose, it is not distinguished between text-based, graphical, and numerical choices. To ensure an unbiased comparison between option categories, the visit duration is normalized to the number of options that the categories consists of. When comparing the visit duration between one correct and the incorrect alternatives, the mean visit duration among the four incorrect alternatives was calculated.
\begin{figure}
     \includegraphics[width=1\linewidth]{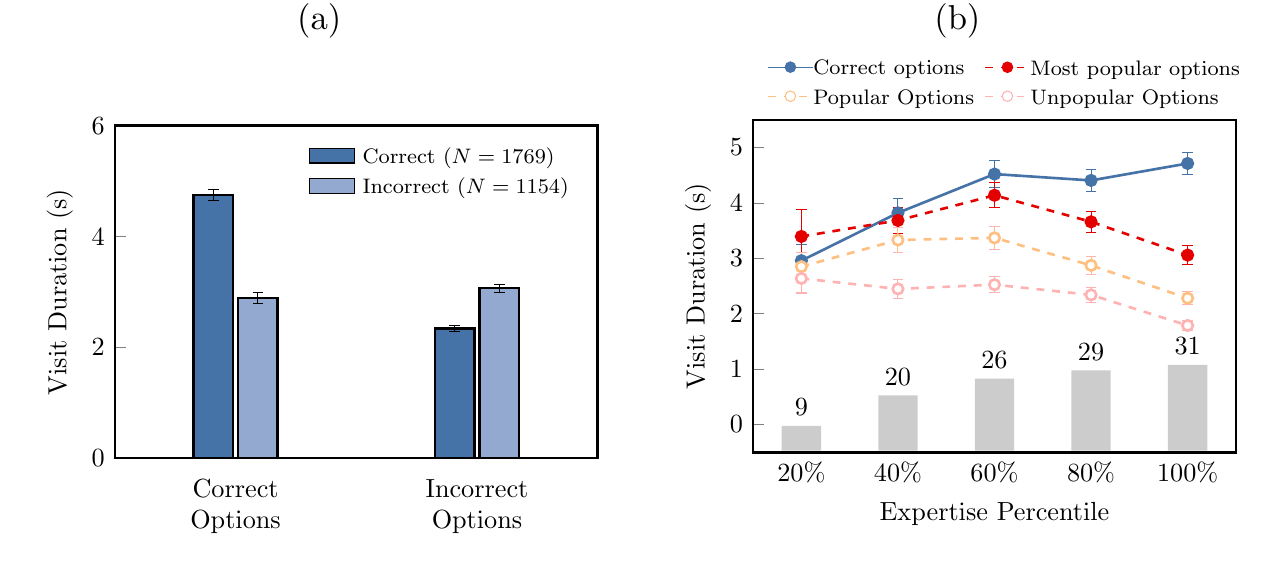}
     \caption{(a) Visit duration on correct and incorrect options for correct and incorrect responses. (b) Visit duration on different option categories as a function of expertise. The histogram in the lower part of the diagram indicates the number of students in each 20\% interval. The analysis includes data of all 26 items. For the definition of the option categories, see Table \ref{tab:options}. The error bars represent the standard errors.}
     \label{fig:categories}
     \end{figure}
     
Figure \ref{fig:categories} (a) shows the visit duration on correct and incorrect options for students who responded correctly and incorrectly. Since every participant fixated on every option type in every item,  the data structure calls for a repeated measure ANOVA ($2 \times 2$) with response accuracy as the between-subject factor and option type as the within-subject factor. We found a significant main effect of option type, $F(1, 2921) = 402, p<10^{-20}$, effect size $d=0.80$. That is, correct options received more attention from students than incorrect options. The interaction between accuracy and option type is also statistically significant with large effect size, $F(1,2919)=540, p<10^{-20}, d=0.95$: Correct responses are associated with more attention on correct options and less attention on incorrect options whereas there is an opposite trend for incorrect answers, confirming the research hypothesis from Sec. \ref{sec:RQ}. This result was obtained using the accuracy and eye-tracking data from all items. By repeating the analysis for every single item, the occurrence and the magnitude of the interaction effect between accuracy and option type reveals whether the hypothesis also holds on the item level. The data is presented in Table \ref{tab:interaction_items}. As can be seen, all interaction effects with one exception (item 12) are significant on the 5\% level and the descriptive data confirm the trend described above, i.e., that the correct options received more attention when students were answering correctly whereas the incorrect responses receive more attention when students were answering incorrectly. The effect sizes range from small (item 20; $d=0.41$) to huge (item 21, $d=5.94$). Since a multitude of 26 statistical tests were conducted on one data set to obtain these results, effects of multiple testing have to be controlled for.  Therefore, we also indicate statistical significance based on a lower threshold defined by the (conservative) Bonferroni correction and find that the interaction effect disappears for the items 13, 20, and 26.
 
     \begin{table}
     \centering\footnotesize
     \caption{Visit duration (VD) in seconds for correct and incorrect options by response accuracy (correct vs. incorrect). Standard errors are given in parentheses. The statistics were obtained from a repeated measure ANOVA and refer to the  interaction  between option type and accuracy on VD.}
     \label{tab:interaction_items}
     \begin{tabular}{cccccccccc} \hline\hline \addlinespace[3pt]
     & \multicolumn{2}{c}{VD correct option} && \multicolumn{2}{c}{VD incorrect options} && \multicolumn{2}{c}{Interaction effect}\\ \cmidrule{2-3} \cmidrule{5-6} \cmidrule{8-9}
     Item & correct & incorrect  && correct & incorrect&& $F$-value$^*$ & effect size \\  \hline\addlinespace[3pt] 
1	&	5.76	(0.47) &	3.39	(0.33) &&	3.55	(0.32) &	4.63	(0.30)&&	45.03$^*$	&	1.61	\\
2	&	1.14	(0.08) &	0.74	(0.14) &&	0.46	(0.04) &	1.00	(0.11)&&	19.12$^*$	&	0.93	\\
3	&	3.71	(0.24) &	3.12	(0.46) &&	3.45	(0.17) &	5.17	(0.42)&&	24.36$^*$	&	1.02	\\
4	&	1.92	(0.26) &	0.75	(0.08) &&	0.82	(0.12) &	0.89	(0.07)&&	50.81$^*$	&	1.86	\\
5	&	1.38	(0.12) &	0.61	(0.11) &&	0.36	(0.04) &	0.71	(0.08)&&	56.62$^*$	&	2.20	\\
6	&	3.44	(0.31) &	1.31	(0.17) &&	1.10	(0.10) &	1.44	(0.12)&&	44.64$^*$	&	1.64	\\
7	&	1.69	(0.16) &	1.00	(0.12) &&	0.60	(0.08) &	1.42	(0.12)&&	59.50$^*$	&	2.16	\\
8	&	6.58	(0.40) &	3.81	(0.49) &&	4.69	(0.27) &	5.41	(0.52)&&	38.41$^*$	&	1.41	\\
9 & 5.85	(0.41) &	2.01	(0.38) &&	3.08	(0.23) &	3.69	(0.37)&&	66.54$^*$	&	2.39	\\

10	&	9.02	(0.62) &	5.63	(0.44) &&	3.30	(0.29) &	3.94	(0.24)&&	48.27$^*$	&	1.70	\\
11	&	5.45	(0.45) &	2.94	(0.42) &&	2.69	(0.20) &	4.24	(0.36)&&	45.71$^*$	&	1.62	\\
12	&	4.31	(0.31) &	4.23	(0.43) &&	4.22	(0.35) &	3.21	(0.36)&&	3.86	&	...	\\
13	&	0.99	(0.07) &	0.61	(0.22) &&	0.20	(0.02) &	0.45	(0.08)&&	6.26	&	0.45	\\
14	&	7.52	(0.47) &	6.08	(1.57) &&	2.40	(0.25) &	4.60	(1.17)&&	10.26$^*$	&	0.60	\\
15	&	10.2	(0.74) &	7.56	(0.96) &&	3.46	(0.50) &	5.38	(0.44)&&	25.63$^*$	&	1.06	\\
16	&	1.29	(0.17) &	0.72	(0.10) &&	0.54	(0.08) &	0.97	(0.09)&&	35.40$^*$	&	1.44	\\
17	&	8.37	(0.50) &	5.18	(0.61) &&	4.05	(0.28) &	4.71	(0.43)&&	34.62$^*$	&	1.30	\\
18	&	2.36	(0.26) &	0.86	(0.10) &&	1.05	(0.21) &	1.19	(0.12)&&	36.72$^*$	&	1.44	\\
19	&	7.81	(0.71) &	3.27	(0.32) &&	2.38	(0.30) &	2.93	(0.21)&&	86.59$^*$	&	3.65	\\
20	&	4.72	(0.62) &	3.75	(0.44) &&	3.02	(0.29) &	3.74	(0.28)&&	5.58	&	0.41	\\
21	&	8.27	(0.37) &	3.02	(0.47) &&	2.63	(0.19) &	3.06	(0.28)&&	115.1$^*$	&	5.94	\\
22	&	4.14	(0.30) &	3.26	(0.38) &&	2.50	(0.19) &	3.06	(0.35)&&	23.42$^*$	&	1.00	\\
23	&	5.26	(0.46) &	2.47	(0.26) &&	2.90	(0.25) &	4.26	(0.34)&&	77.94$^*$	&	2.95	\\
24	&	5.20	(0.39) &	2.91	(0.30) &&	3.21	(0.22) &	3.67	(0.28)&&	47.65$^*$	&	1.70	\\
25	&	4.19	(0.26) &	1.87	(0.18) &&	1.74	(0.10) &	1.99	(0.18)&&	38.90$^*$	&	1.46	\\
26	&	3.35	(0.29) &	2.89	(0.52) &&	2.31	(0.21) &	2.98	(0.44)&&	10.13	&	0.59	\\\hline\hline \addlinespace[3pt]
\multicolumn{9}{l}{With exception of item 12, all effects are significant on the $p=0.05$ level. $^*p<0.05/26\approx0.002$}
     \end{tabular}
     \end{table}
     
For a more detailed analysis of the answer choices, the distractor analysis from the original data was used \cite{Zavala}. Based on the frequency of answer choices published by Zavala \textit{et al.}, three categories of incorrect alternatives have been defined (most popular, popular, and unpopular; see Sec. \ref{sec:materials} and Table \ref{tab:options}) and the mean visit duration on each category was determined for every person and every item. The visit duration on every option category is shown in Fig. \ref{fig:categories} (b) as a function of the students' expertise (that was determined by the total test score). A repeated measures ANOVA ($4 \times 5$) was conducted with the option category as the within-subject variable and the expertise level as the between-subject variable. A significant main effect of the factor option type was found, $F(3, 4734)=110.4, p<10^{-20}, d=0.55$, indicating that the attention was not equally distributed among the different option types.  Post-hoc analyses with pair-wise comparisons revealed that the correct responses received significantly more attention than all other types of answer choices with effect sizes between $d=0.14$ and $d=0.46$. The unpopular choices received the least attention ($d=0.18-0.46$), and the most popular and popular incorrect options are in between (separated from each other with effect size $d=0.16$). Furthermore, a significant interaction effect with small effect size was confirmed, $F(12, 4734)=8.1, p<10^{-15}, d=0.27$. The interaction between students' expertise level and option type is most pronounced for the highest expertise levels. When transitioning from the 80\% to the 100\% expertise level, the students focus more often on the correct options and their visual attention on all other options decreases. As can be seen from Fig. \ref{fig:categories} (b), the difference between the students' attention on the most popular incorrect options and the correct options increases from low to high expertise levels.

\section{Discussion and conclusion}
In this study, we used eye tracking to capture the students' visual attention while they solved the test of understanding graphs in kinematics (TUG-K ) during a computer-based assessment scenario. Besides choosing the correct alternative among research-based distractors, the students were required to judge their response confidence for each question. Even though all students were exposed to the subject at school, the mean test test score of 59\% does not indicate mastery of all concepts in the test, indicating that more emphasis should be put on graph interpretation. Results from recent studies also highlight the importance of an instructional adjustment towards a more graphical-based education \cite{Susac, Klein2019}.  Especially the items 1, 4, 10, 16, and 19 which deal with the area concept were among the hardest for the students, confirming previous findings about students' difficulties with interpretation of area under a curve \cite{Bei94, Bei96, Plani13}.

Overall, the students provided correct answers with higher confidence ratings in comparison to when they gave incorrect answers. Thus, the students' ability and their confidence were highly intercorrelated. However, some items show below-average item difficulty but above average confidence scores, indicating that the correlation between response accuracy and response confidence varies among the items. For instance, students were quite confident (70\%) when responding to item 4 (calculating the area under a curve) but they were correct only in 38\% of the cases. The strongest distractor (E, chosen by 36\%  of students) reflects the popular error of determining the area of a rectangle instead of a triangle.  Therefore, the participants were able to select an incorrect alternative that reflected their flawed thinking process, resulting in a high response confidence; an observation that was also examined elsewhere \cite{KIRC}. It was not in the scope of this paper but the analysis of confidence scores besides response accuracy might reveal more information about student misconceptions, and we encourage researchers to add confidence scales to their assessment. 

From the eye-tracking data, we found that the TUG-K took about 20 min to complete on average. We found a significant correlation between the time spent on completing the items and the confidence index with longer visit durations indicating low confidence. This result was confirmed by analysing the time spent on the questions and option choices with respect to the students confidence ratings on single test items. The effect was more pronounced for the options ($d=0.34$) than for the questions ($d=0.20$) and the visit duration on the options provided a good  discriminator between low, intermediate and high confidence levels. The magnitudes of effect sizes are similar to those reported in a previous study comparing visit durations and confidence ratings \cite{Kuechemann}. Students responding with high confidence presumably think to know what answer choice to look  for  (in case of number choices) or which features of the graphs are relevant to solve the questions (in case of graph choices) so it takes less time to evaluate the options. On the contrary, students with low response confidence might take more options into consideration, comparing them and therefore need more time to select an alternative. The data also showed a very small interaction between confidence and accuracy on the time spent on the question:  High confident students spent less time on the question when answering correct compared to students that answer incorrect. For students with low confidence it is vice versa and further qualitative research is required to explain this result. 

There was no correlation between time spent on the test and the students' performance in general. There was also no difference in the visit duration between correct and incorrect responses, neither regarding the item stems (questions) nor the options, confirming the result obtained from Han \textit{et al.} in the context of the FCI \cite{Han}. However, similar to Han \textit{et al.}, when the options are split into correct and incorrect choices, the attention on these choices discriminates the correct from the incorrect answers in terms of an interaction effect: Incorrect responses are associated with longer visit durations on incorrect options and less time spent  on correct options while correct responses show the opposite trend. This confirms a general trend of answering multiple choice questions found by Tsai \textit{et al.}; while solving a multiple-choice science problem, students pay more attention to chosen options than rejected alternatives, and spent more time inspecting relevant factors than irrelevant ones \cite{Tsai}. Given that the interaction effect occurs for (almost) every single item of the TUG-K and on the test level (26 items), we can consider the result obtained in this study as a broad generalization of Tsai  \textit{et al.} who investigated a rather small sample of six students and only one item. The effect size ranges from medium to huge on the individual item level and the average effect size (considering all items) is large, pointing towards the practical importance of the effect. 
However, there are very few exceptions with no effect at all (item 12) or small effect sizes (items 13, 20, and 26) that should be mentioned: Three of these items (12, 20, 26) require the selection of a graph from a textual description. The options show five graphs (I, II, ..., V) and the test taker must judge which of the graphs suit a condition (e.g., that shows a motion with constant speed). The final options are presented in a complex multiple choice format, e.g. ``(a) graphs I, III and V", ``(b) graphs I and III", etc. Therefore, incorrect answers---for instance ``(a) graphs I, III and IV"---involve a subset of correct graphs---for instance I and III---and therefore the analysis procedure cannot discriminate between the visit duration on correct and incorrect options by correct and incorrect responses. For future analysis, we advocate to treat these items separately. Additionally, we encourage test developers to change the response format from complex multiple choice to single choice to have a consistent format.

Finally, the incorrect option choices were divided into three categories based on the empirical data about selection frequency obtained in the original study \cite{Zavala}. It was found that unpopular options receive the least attention among all levels of expertise. The higher the expertise level of the students, the more attention is allocated on the correct options and the less attention is allocated on the most popular options. Even in the highest levels of expertise, the most popular incorrect options, that were designed to address popular misconceptions and learning difficulties with graphs,  receive more attention that the other incorrect options. Expert students (as indicated by high test scores) who shifted their attention to the correct choices still keep a higher level of attention to the popular options, indicating conceptual mixing \cite{Han}. From an assessment point of view, this result provides evidence for the test validity at the behavioural level. The choices were well designed to reflect typical flawed thinking processes that students have to encounter during problem solving. Future work could focus on specific alternative conceptions and student errors while taking the TUG-K and whether they are related to certain eye-gaze patterns. 

In summary, the study showed that eye tracking can make unique contributions to the validation of concept inventories on a behavioural level without using interview or survey data. While simple time measures (visit durations on the question or options) are well suited to discriminate between different confidence levels of the test takers, they do not discriminate the correct from the incorrect performers for this type of questions. A more fine-grained partition of option choices---based on educational considerations or empirical data---is required to relate the students' accuracy to eye-tracking measures.

\section*{Appendix}
The categorization of answer options is shown in Table \ref{tab:options} and a short overview on the items is given in Table \ref{tab:content}.
\begin{table}[b!]
\centering \footnotesize
\caption{Categorization of options based on the empirical results from the original work by  Zavala \textit{et al.} \cite{Zavala}}
\label{tab:options}
\begin{tabular}{ccccc} \hline \hline
Item	&	correct	&	most popular	&	popular	&	unpopular	\\\hline
1	&	A	&	D	&	E	&	B,C	\\
2	&	E	&	C	&	B	&	A,D	\\
3	&	D	&	C	&	...	&	A,B,E	\\
4	&	D	&	E	&	B,C	&	A	\\
5	&	C	&	D	&	...	&	A,B	\\
6	&	B	&	A	&	C,D,E	&	...	\\
7	&	A	&	C	&	B,D	&	E	\\
8	&	D	&	B,C	&	E	&	A	\\
9	&	E	&	B	&	...	&	A,C,D	\\
10	&	A	&	B	&	...	&	C,D,E	\\
11	&	D	&	B	&	A,C	&	E	\\
12$^*$	&	I, III	&	V	&	I	&	 II	\\
13	&	A	&	D	&	B	&	C,E	\\
14	&	B	&	D	&	E	&	A,C	\\
15	&	A	&	E	&	C,D	&	B	\\
16	&	D	&	B	&	A,C,E	&...		\\
17	&	A	&	D	&	...	&	B,C,E	\\
18	&	A	&	D	&	B,C,E	&	...	\\
19	&	B	&	D	&	C	&	A,E	\\
20$^*$	&	 V	&	 II, IV	&	...	&	 I, III	\\
21	&	B	&	D	&	E	&	A,C	\\
22$^*$	&	 II, V	&	...	&	 III	&	 I, IV	\\
23	&	B	&	E	&	C,D	&	A	\\
24	&	A	&	B	&	C	&	D,E	\\
25	&	C	&	D	&	...	&	A,B,E	\\
26$^*$	&	 III, V	&	 II	&	...	&	 I, IV	\\
\hline \hline
\multicolumn{5}{c}{$^*$Option choices A--E are related to graph choices I--V}
\end{tabular}
\end{table}

\begin{table}[b!] \footnotesize
\centering
\caption{Overview on the 26 TUG-K items} \label{tab:content}
\begin{tabular}{p{.3\linewidth}lp{.4\linewidth}} \hline \hline \addlinespace[3pt]
Item type & Items & Description \\ \hline \addlinespace[3pt]
Text $\rightarrow$ Graphs & 1, 23 & Identify a graph with greatest change in a variable \\
	&9,12,20,22,26 & Select a graph from a textual description\\\addlinespace[5pt]
Text + Graph $\rightarrow$ Graphs & 11, 14, 15, 21 & Select corresponding graph from a graph\\\addlinespace[5pt]
Text + Graph $\rightarrow$ Text& 3, 8, 17, 24, 25 & Select a textual description from a graph \\
	 &10, 19 & Establish the procedure to determine the change of a variable \\
Text + Graph $\rightarrow$ Values & 2, 5, 6, 7, 13, 18 & Evaluate the slope of a graph\\
&	4, 16 & Evaluate the area under the curve \\
\hline \hline
\end{tabular}
\end{table}


\section*{References}
\begin{thebibliography}{20}

\bibitem{Bei94} Beichner R J 1994  Testing student interpretation of kinematics graphs \textit{American journal of Physics} \textbf{62}  750-762

\bibitem{Perez10} Perez-Goytia N, Dominguez A and Zavala G 2010 Understanding
and interpreting calculus graphs: Refining an instrument \textit{AIP Conference Proceedings} \textbf{1289} 249 

\bibitem{MS13} Maries A and Singh C 2013 Exploring one aspect of pedagogical
content knowledge of teaching assistants using the test of understanding graphs in kinematics \textit{Physical Review Special Topics Physics Education Research} \textbf{9} 020120 

\bibitem{Gurel15} Gurel D K,  Eryılmaz A and McDermott L C 2015 A Review and Comparison of Diagnostic Instruments to Identify Students' Misconceptions in Science \textit{Eurasia Journal of Mathematics, Science \& Technology Education} \textbf{11}

\bibitem{Bei96} Beichner R J 1996 The impact of video motion analysis on kinematics graph interpretation skills \textit{American Journal of physics} \textbf{64} 1272--1277

\bibitem{Bektasli12} Bektasli B and White A L 2012 The Relationships between Logical Thinking, Gender, and Kinematics Graph Interpretation Skills \textit{Eurasian Journal of Educational Research} \textbf{48} 1--19

\bibitem{Lichtenberger} Lichtenberger A, Wagner C, Hofer S I, Stern E and Vaterlaus A 2017  Validation and structural analysis of the kinematics concept test. \textit{Physical Review Physics Education Research} \textbf{13} 010115

\bibitem{KIRC} Klein P, M\"{u}ller A and Kuhn J 2017 Assessment of representational competence in kinematics \textit{Physical Review Physics Education Research} \textbf{13} 010132

\bibitem{Zavala} Zavala G, Tejeda S, Barniol P and Beichner R J 2017 Modifying the test of understanding graphs in kinematics \textit{Physical Review Physics Education Research} \textbf{13} 020111

\bibitem{Lindell} Lindell R S, Peak E and Foster T M 2007 Are they all created equal? A comparison of different concept inventory development methodologies \textit{AIP conference proceedings} \textbf{883} 14--17

\bibitem{Adams} Adams W K and Wieman C E 2011 Development and validation of instruments to measure learning of expert-like thinking \textit{International Journal of Science Education} \textbf{33} 1289--1312

\bibitem{Polking07} Polkinghorne D E 2007 Validity issues in narrative research \textit{Qualitative inquiry} \textbf{13} 471--486

\bibitem{Kozhevnikov} Kozhevnikov M, Motes M A and Hegarty M 2007 Spatial visualization in physics problem solving \textit{Cognitive Science} 31 549--579

\bibitem{Madsen} Madsen A M, Larson A M, Loschky L C and Rebello N S 2012 Differences in visual attention between those who correctly and incorrectly answer physics problems \textit{Physical Review Special Topics Physics Education Research} \textbf{8} 010122

\bibitem{Madsen2} Madsen A, Rouinfar A, Larson A M, Loschky L C and Rebello N S 2013 Can short duration visual cues influence students' reasoning and eye movements in physics problems?
\textit{Physical Review Special Topics Physics Education Research} \textbf{9} 020104

\bibitem{Susac} Susac A, Bubic A, Kazotti E, Planinic M and Palmovic M 2018 Student understanding of graph slope and area under a graph: A comparison of physics and nonphysics students  \textit{Physical Review Physics Education Research} \textbf{14} 020109

\bibitem{Klein2019} Klein P, K\"{u}chemann S, Br\"{u}ckner S, Zlatkin-Troitschanskaia O and Kuhn J 2019 Student understanding of graph slope and area under a curve: A replication study comparing first-year physics and economics students \textit{Physical Review Physics Education Research} (Accepted 26 July 2019)


\bibitem{Martina1} Kekule M 2015 Students' different approaches to solving problems from kinematics in respect of good and poor performance, \textit{Proceedings of the International Conference on Contemporary Issues in Education}  126--34

\bibitem{Martina2} Kekule M 2014 Students' approaches when dealing with kinematics graphs explored by eye-tracking research method, \textit{Proceedings of the frontiers in mathematics and science education research conference}  108--117

\bibitem{Han} Han J, Chen L, Fu Z, Fritchman J and Bao L 2017 Eye-tracking of visual attention in web-based assessment using the Force Concept Inventory \textit{European Journal of Physics} \textbf{38} 045702

\bibitem{Aslanides2013} Aslanides J S and Savage C M 2013 Relativity concept inventory: Development, analysis, and results \textit{Physical Review Special Topics Physics Education Research} \textbf{9}, 010118  

\bibitem{Hasan} Hasan S, Bagayoko D and Kelley E L Misconceptions and the certainty of response index 1999 \textit{Physics Education} \textbf{34} 294--299 

\bibitem{Lindsey} Lindsey B A and Nagel M L 2015 Do students know what they know? Exploring the accuracy of students' self-assessments \textit{Physical Review Special Topics Physics Education Research} \textbf{11} 020103

\bibitem{Dowd} Dowd J E, Araujo I and Mazur E 2015 Making sense of confusion: Relating performance, confidence, and self-efficacy to expressions of confusion in an introductory physics class \textit{Physical Review Special Topics Physics Education Research} \textbf{11} 010107 

\bibitem{Kuechemann} K\"{u}chemann S, Klein P, Fouckhardt H, Gr\"{o}ber S and Kuhn J 2019 Improving studens' understanding of rotating frames of reference using videos from different perspectives, arXiv:1902.10216 [physics.ed-ph]

\bibitem{Geg11} Gegenfurtner A, Lehtinen E and S\"{a}lj\"{o} R 2011 Expertise Differences in the Comprehension of Visualizations: a Meta-Analysis of Eye-Tracking Research in Professional Domains \textit{Educational Psychology Review} \textbf{23} 523--552

\bibitem{physport} https://www.physport.org

\bibitem{tobii} More specifications can be found on the product website https://www.tobiipro.com

\bibitem{salvucci2000identifying}  Salvucci D D and  Goldberg J H 2000 Identifying fixations and saccades in eye-tracking protocols \textit{Proceedings of the 2000 symposium on Eye tracking research \& applications} 71--78

\bibitem{Cohen} Cohen J 1992 A power primer \textit{Psychological Bulletin} \textbf{112}  155--159

\bibitem{Sawil09} Sawilowsky S 2009 New effect size rules of thumb \textit{Journal of Modern Applied Statistical Methods} \textbf{8}  467--474

\bibitem{DB09} Ding L and Beichner R 2009 Approaches to data analysis
of multiple-choice questions \textit{Physical Review Special Topics Physics Education Research} \textbf{5} 020103

\bibitem{Plani13} Planinic M, Ivanjek L, Susac A and Milin-Sipus Z 2013 Comparison of university students' understanding of graphs in different contexts \textit{Physical Review Special Topics Physics Education Research} \textbf{9} 020103 

\bibitem{Tsai} Tsai M J, Hou H T, Lai M L, Liu W Y and Yang F Y 2012 Visual attention for solving multiple-choice science problem: An eye-tracking analysis \textit{Computers \& Education} \textbf{58} 375--385

\end{thebibliography}
\end{document}